\documentclass[reprint,amsmath,amssymb,aps,floatfix,superscriptaddress]{revtex4-1}
\usepackage[utf8]{inputenc}
\usepackage[T1]{fontenc}
\usepackage{microtype}
\usepackage{physics}
\usepackage{lmodern}
\usepackage{graphicx}
\usepackage{textcomp}
\usepackage[hidelinks, bookmarks=false]{hyperref}
\usepackage[capitalise]{cleveref}
\usepackage{siunitx}

\begin{document}
\title{Improved success probability with greater circuit depth for the quantum approximate optimization algorithm}
\author{Andreas Bengtsson}
\author{Pontus Vikstål}
\author{Christopher Warren}
\affiliation{Microtechnology and Nanoscience, Chalmers University of Technology, SE-412 96, Göteborg, Sweden}
\author{Marika Svensson}
\affiliation{Computer Science and Engineering, Chalmers University of Technology, SE-412 96, Göteborg, Sweden}
\affiliation{Jeppesen, SE-411 03, Göteborg, Sweden}

\author{Xiu Gu}
\author{Anton Frisk Kockum}
\author{Philip Krantz}
\author{Christian Križan}
\author{Daryoush Shiri}
\author{Ida-Maria Svensson}
\author{Giovanna Tancredi}
\author{Göran Johansson}
\author{Per Delsing}
\author{Giulia Ferrini}
\author{Jonas Bylander}
\email{bylander@chalmers.se}
\affiliation{Microtechnology and Nanoscience, Chalmers University of Technology, SE-412 96, Göteborg, Sweden}

\date{\today}
\begin{abstract}

Present-day, noisy, small or intermediate-scale quantum processors---although far from fault-tolerant---support the execution of heuristic quantum algorithms, which might enable a quantum advantage, for example, when applied to combinatorial optimization problems.
On small-scale quantum processors, validations of such algorithms serve as important technology demonstrators. 
We implement the quantum approximate optimization algorithm (QAOA) on our hardware platform, consisting of two superconducting transmon qubits and one parametrically modulated coupler.
We solve small instances of the NP-complete exact-cover problem, with 96.6\% success probability, by iterating the algorithm up to level two.

\end{abstract}
\maketitle

\section{Introduction}
Quantum computing promises exponential computational speedup in a number of fields, such as cryptography, quantum simulation, and linear algebra \cite{montanaro2016quantum}. Even though a large, fault-tolerant quantum computer is still many years away, impressive progress has been made over the last decade using superconducting circuits \cite{wendin2017quantum, GU20171, kjaergaard2019superconducting}, leading to the noisy intermediate-scale quantum (NISQ) era \cite{Perskillnisq}. It was predicted that NISQ devices should allow for ``quantum supremacy'' \cite{boixo2018characterizing}, that is, solving a problem that is intractable on a classical computer in a reasonable time. This was recently demonstrated on a 53-qubit processor by sampling the output distributions of random circuits~\cite{arute2019quantum}. 

Two of the most prominent NISQ algorithms are the quantum approximate optimization algorithm (QAOA), for combinatorial optimization problems \cite{farhi2014quantum, otterbach2017unsupervised, pagano2019quantum}, and the variational quantum eigensolver (VQE), for the calculation of molecular energies \cite{peruzzo2014variational, o2016scalable, kandala2017hardware}. QAOA is a heuristic algorithm that could bring a polynomial speedup to the solution of specific problems encoded in a quantum Hamiltonian \cite{jiang2017near,niu2019optimizing}. Moreover, QAOA should produce output distributions that cannot be efficiently calculated on a classical computer \cite{farhi2016quantum}.

QAOA is a hybrid algorithm, as it is executed on both a classical and a quantum computer. The quantum part consists of a circuit with $p$ levels, where better approximations to the solution of the encoded problem are generally achieved with higher $p$. In this work, we report on using our superconducting quantum processor to demonstrate QAOA with up to $p=2$, enabled by adequately high gate fidelities. We solve small toy instances of the NP-complete exact-cover problem with 96.6\% success probability. For $p>1$, the QAOA solution cannot be efficiently calculated on a classical computer, as the computational complexity scales doubly exponentially in $p$ \cite{farhi2014quantum}. 

Our interest in solving the exact-cover problem originates from its use in many real-world applications, for instance, the exact-cover problem can provide feasible solutions to airline planning problems such as tail assignment \cite{Gronkvist}. Currently, this is solved by well-developed optimization techniques in combination with heuristics. By leveraging heuristic quantum algorithms such as QAOA, the current approach can be augmented and might provide high-quality solutions while reducing the running time. Applying QAOA to instances of the exact-cover problem extracted from real-world data in the context of the tail assignment has been numerically studied with 25 qubits, corresponding to 25 routes and 278 flights \cite{Vikstal2019}. Other quantum algorithms for solving the exact-cover problem, specifically quantum annealing, have been considered in Refs.~\cite{farhi2001quantum, wang2016ultrafast, Grass2019quantum}.

\begin{figure}
    \centering
    \includegraphics[width=1\linewidth]{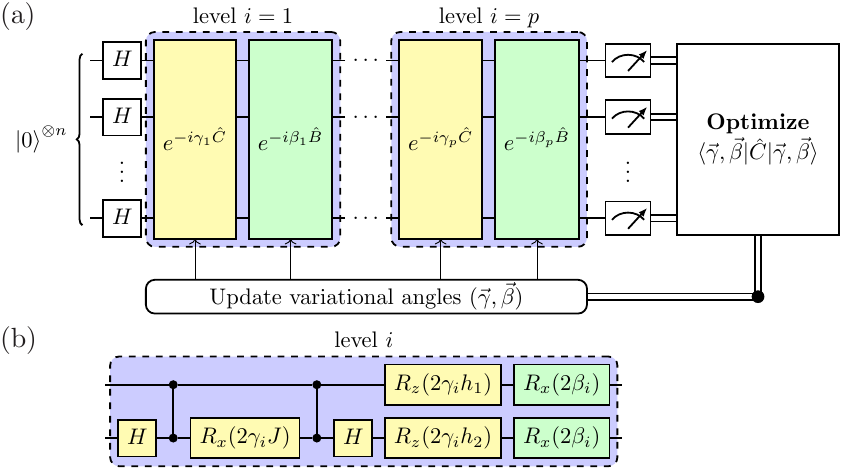}
    \caption{(a) The quantum approximate optimization algorithm (QAOA) for a problem specified by the Ising Hamiltonian $\hat{C}$. An alternating sequence of two Hamiltonians ($\hat{C}$ and $\hat{B}$) is applied to an equal superposition of $n$ qubits. After measurement of the qubit states, a cost is calculated, which a classical optimization algorithm minimizes by varying the angles $\vec{\gamma},\vec{\beta}$.
    (b) Our implementation of one QAOA level with $n=2$ using controlled-phase and single-qubit gates. The background color of each gate identifies which part in (a) it implements. 
    }
    \label{fig:theory}
\end{figure}

\section{QAOA}

All NP-complete problems can be formulated in terms of finding the ground state of an Ising Hamiltonian \cite{lucas2014ising}. QAOA aims at finding this state by applying two non-commuting Hamiltonians, $\hat{B}$ and $\hat{C}$, in an alternating sequence (with length $p$) to an equal superposition state of $n$ qubits [visualized in \cref{fig:theory}(a)],
\begin{equation}
    \label{eq:state}
    \ket*{\vec{\gamma},\vec{\beta}} = \prod_{i=1}^p \bigg[e^{-i\beta_i \hat{B}} e^{-i\gamma_i \hat{C}}\bigg]\bigg(\frac{\ket{0}+\ket{1}}{2}\bigg)^{\otimes n},
\end{equation}
where $\gamma_i$ and $\beta_i$ are (real) variational angles. 
The first Hamiltonian in the sequence is the Ising (cost) Hamiltonian specifying the problem,
\begin{equation}
    \label{eq:H}
    \hat{C} = \sum_{i=1}^n h_i\hat{\sigma}_i^z + \sum_{i<j} J_{ij}\hat{\sigma}_i^z\hat{\sigma}_j^z, 
\end{equation}
and the second is a transverse field (mixing) Hamiltonian defined by
\begin{equation}
    \hat{B} = \sum_{i=1}^n \hat{\sigma}_i^x,
\end{equation}
where $h_i$ and $J_{ij}$ are real coefficients, and $\hat{\sigma}_i^{x (z)}$ are the Pauli X (Z) operators applied to the $i^{\text{th}}$ qubit.

The ground state of \cref{eq:H} corresponds to the lowest-energy state. We therefore define the energy expectation value of \cref{eq:state} as a cost function
\begin{equation}
    \label{eq:F}
    F(\vec{\gamma}, \vec{\beta}) = \expval*{\hat{C}}{\vec{\gamma},\vec{\beta}} =
    \sum_{i=1}^n h_i\langle\hat{\sigma}_i^z\rangle  + \sum_{i<j} J_{ij}\langle\hat{\sigma}_i^z\hat{\sigma}_j^z\rangle.
\end{equation}
This cost function is evaluated by repeatedly preparing and measuring $\ket*{\vec{\gamma},\vec{\beta}}$ on a quantum processor. To find the state that minimizes \cref{eq:F}, a classical optimizer is used to find the optimal variational angles $\vec{\gamma}^*,\vec{\beta}^*$. For a high enough $p$, $\ket*{\vec{\gamma}^*,\vec{\beta}^*}$ is equal to the ground state of $\hat{C}$ and hence yields the answer to the optimization problem \cite{farhi2014quantum}. However, for algorithms executed on real hardware without error correction, noise will inevitably limit the circuit depth, implying that there is a trade-off between algorithmic errors (too low $p$) and gate errors (too high $p$). 
Note that, in order to find the solution to the optimization problem, it is not necessary for $\ket*{\vec{\gamma}^*,\vec{\beta}^*}$ to be equal to the ground state: as long as the ground-state probability is high enough, the quantum processor can be used to generate a shortlist of potential solutions that can be checked efficiently (in polynomial time) on a classical computer. For instance, even if the success probability of measuring the ground state is only 5\%, we could measure 100 instances and still attain a probability greater than 99\% of finding the correct state. Moreover, the angles $\vec{\gamma}^*,\vec{\beta}^*$ themselves are not interesting, as long as they yield the lowest-energy state. This gives some robustness against coherent gate errors, since any over- or under-rotations can be compensated for by a change in the variational angles \cite{o2016scalable}.

We apply QAOA to the exact-cover problem, which reads: given a set $X$ and several subsets $S_i$ containing parts of $X$, which combination of subsets include all elements of $X$ just once? Mathematically speaking, this combination of subsets should be disjoint, and their union should be $X$. This problem can be mapped onto an Ising Hamiltonian, where the number of spins equals the number of subsets, while the size of $X$ can be arbitrary.

Let us consider $n=2$, for which the two-spin Ising Hamiltonian is 
\begin{equation}
    \label{eq:H2}
    \hat{C} = h_1\hat{\sigma}_1^z + h_2\hat{\sigma}_2^z +J\hat{\sigma}_1^z\hat{\sigma}_2^z.
\end{equation}
The exact-cover problem is mapped onto this Hamiltonian by choosing $h_i$ and $J$ as follows \cite{choi2010adiabatic}:
\begin{align}
\label{eq:crit}
    J &> \mathrm{min}(c_1, c_2), \\
    h_1 &= J - 2 c_1, \nonumber\\
    h_2 &= J - 2 c_2, \nonumber
\end{align}
where $c_i$ is the number of elements in subset $i$, and $J>0$ if the two subsets share at least one element.
We are free to choose $J$, as long as it fulfills the criterion in \cref{eq:crit}. For example, consider $X=\{x_1, x_2\}$ and two subsets $S_1=\{x_1, x_2\}$ and $S_2=\{x_1\}$. This gives $c_1=2$ and $c_2=1$, and we could choose $J = 2$, yielding $h_1=-2$ and $h_2=0$.  It is easy to check that the corresponding ground state is $\ket{10}$ (i.e., $S_1$ is the solution). Finally, we normalize $J$ and $h_i$ such that the Ising Hamiltonian has integer eigenvalues, allowing us to restrict $\gamma_i$ and $\beta_i$ to the interval $[0, \pi[$.

For two subsets, four different problems exist, which all yield different sets of $h_i$ and $J$. These are summarized in \cref{tab:problems}. Problem A is the example given above; it is the most interesting, as the other problems are trivial. Problems B and C are trivial since they do not contain any qubit-qubit interaction ($J=0$). Problem D is also trivial since both subsets are equal. Additionally, the ground states are degenerate for problems B and D.

\begin{table}
\centering
\caption{The four different exact-cover problems available with two subsets, and their solutions and respective sets of coefficients in the Ising Hamiltonian $\hat{C} = h_1\hat{\sigma}_1^z + h_2\hat{\sigma}_1^z +J\hat{\sigma}_1^z\hat{\sigma}_2^z$.
}
\label{tab:problems}
\begin{ruledtabular}
\begin{tabular}{ p{0.8cm} p{2.4cm} | p{0.8cm} p{0.8cm} p{0.8cm} p{1.9cm}}
  \# & Subsets & $h_1$ & $h_2$ & $J$ & Solution  \\
  \hline
  A & $S_1=\{x_1, x_2\}$,\newline $S_2=\{x_1\}$ & -1/2 & 0 & 1/2 & $\ket{10}$ \\
  B & $S_1=\{x_1, x_2\}$,\newline $S_2=\{\}$ & -1 & 0 & 0 & $\ket{10}$ or $\ket{11}$ \\
  C & $S_1=\{x_1\}$,\newline $S_2=\{x_2\}$ & -1/2 & -1/2 & 0 & $\ket{11}$ \\
  D & $S_1=\{x_1, x_2\}$,\newline $S_2=\{x_1, x_2\}$ & 0 & 0 & 1 & $\ket{10}$ or $\ket{01}$ \\
\end{tabular}
\end{ruledtabular}
\end{table}

\section{Realization on quantum hardware}

We implement \cref{eq:state} on our quantum processor using the circuit in \cref{fig:theory}(b). The circuit can be somewhat compiled by simple identities (e.g., two Hadamard gates equal identity). We stress that our implementation of QAOA is scalable in that we do not use any exponentially costly pre-compilation (e.g., calculating the final circuit unitary and using Cartan decomposition to minimize the number of two-qubit gates).  

Our quantum processor is fabricated using the same processes as in Ref.~\cite{burnett2019decoherence} and consists of two fixed-frequency transmon qubits with individual control and readout. Both qubits are coupled to a common frequency-tunable coupler used to mediate a controlled-phase gate (CZ) between the qubits. The CZ gate is realized by a full coherent oscillation between the $\ket{11}$ and $\ket{02}$ states. The interaction is achieved by parametrically modulating the resonant frequency of the coupler at a frequency close to the difference frequency between the $\ket{0} - \ket{1}$ and $\ket{1} - \ket{2}$ transitions of qubit 1 and 2, respectively \cite{McKayUniversal, caldwell2018parametrically}. We have benchmarked such a gate on the same device during the same cooldown to above 99\%; however, the benchmark performed closest in time to the experiments presented here showed a fidelity of 98.6\%. These kinds of fidelity fluctuations might be related to fluctuations in the qubits' coherence times \cite{burnett2019decoherence}.
Single-qubit X rotations are driven by microwave pulses at the qubit transition frequencies with fidelities of 99.86\% and 99.93\% for the respective qubits. Z-rotations are implemented in software as a shift in drive phase and thus have unity fidelity \cite{mckay2017efficient}. All the reported gate fidelities were measured by (interleaved) randomized benchmarking \cite{corcoles2013process}. More experimental details, a measurement setup along with a device schematic, and benchmarking results are found in \cref{app:setup} and \cref{app:char}. 

\section{Applying QAOA to four problems}
For $p=1$, we apply a simple grid ($61 \times 61$) search of $\beta_1,\gamma_1 \in [0, \pi[$ while recording 5000 measurements of each qubit. From these, we calculate $\langle \sigma_i^z \rangle$, $\langle \sigma_1^z\sigma_2^z \rangle$, the cost function $F$, and the occupation probability for each of the four possible states, while accounting for the limited, but calibrated, readout fidelity (\SI{86}{\percent} and \SI{95}{\percent} for the two qubits). By collecting sufficiently many samples, the statistical error on the estimated quantities can be made small. 

\begin{figure}
    \centering
    \includegraphics[width=1\columnwidth]{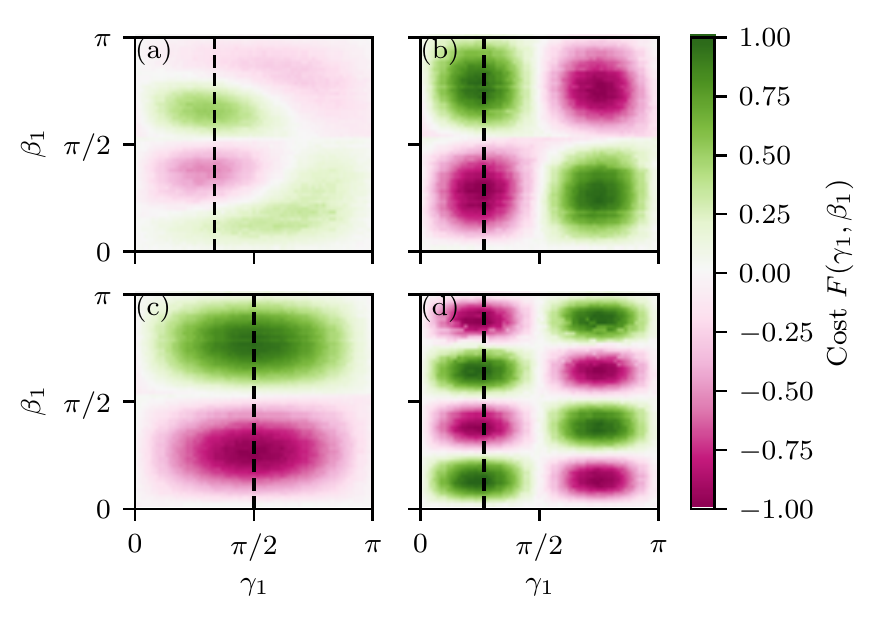}
    \caption{Cost functions $F(\vec{\gamma},\vec{\beta})$ for QAOA applied to four instances of the exact-cover problem with $p=1$ and $n=2$. (a-d) correspond to problems A-D in \cref{tab:problems}. Each experimental data point is evaluated from the average of 5000 measurements on our quantum processor. The dashed lines indicate the positions of the linecuts in \cref{fig:linecuts}.}
    \label{fig:landscapes}
\end{figure}

The grid search allows us to explore the shape of the optimization landscape, which may bring important understanding in the difficulty of finding global minima for black-box optimizers. 
In \cref{fig:landscapes}, we show measured cost functions for the four problems in \cref{tab:problems}. Due to the normalization of $h_i$ and $J$, the ground state for each problem corresponds to $F=-1$. 
In \cref{fig:landscapes}(a), the cost function for problem A never reaches below $-0.5$. To achieve costs approaching -1, additional levels ($p>1$) are needed. Moreover, the existence of a local minimum around $\gamma_1\approx\beta_1\approx 3\pi/4$ could cause difficulties for optimizers trying to find the global minimum.
For problems B-D [\cref{fig:landscapes}(b-d)], we see clear minima where $F \approx -1$, indicating that we have found the optimal variational angles $\ket*{\vec{\gamma}^*, \vec{\beta}^*}$ corresponding to the ground state.

\begin{figure}
    \centering
    \includegraphics[width=1\columnwidth]{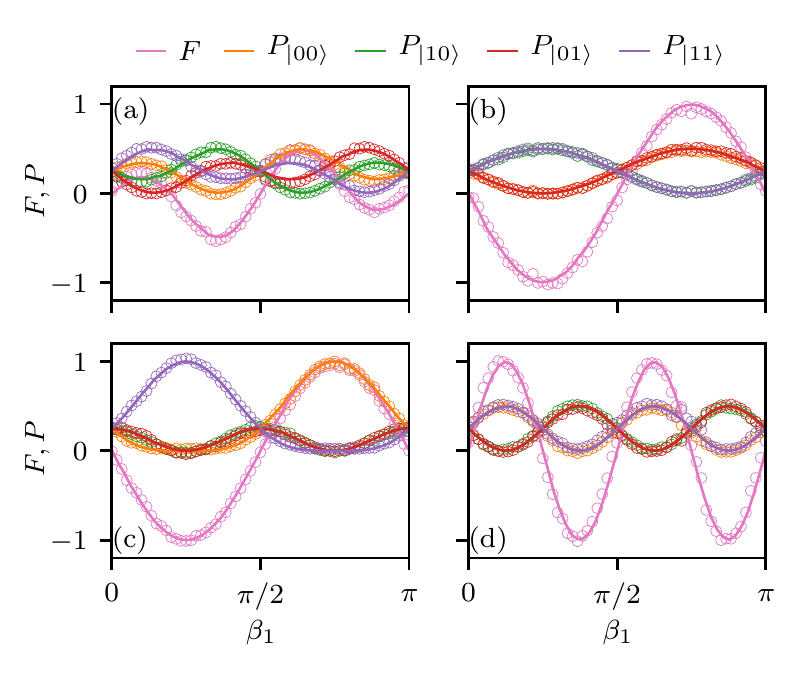}
    \caption{A comparison between experiment (open circles) and theory (solid lines) for four exact-cover problems using QAOA with $p=1$. Each color (given at the top) corresponds to either a state probability or the value of the cost function $F$. The four panels (a-d) correspond to the four problems (A-D) in \cref{tab:problems}. The linecuts are taken at the vertical dashed lines in \cref{fig:landscapes}. The theory curves are calculated assuming an ideal quantum processor, whereas each experimental data point is derived from the average of 5000 measurements on our quantum processor.}
    \label{fig:linecuts}
\end{figure}

In \cref{fig:linecuts}, we take linecuts along the dashed lines in \cref{fig:landscapes} and benchmark our measured cost functions and state probabilities against those of an ideal quantum computer without any noise. We see excellent agreement between measurement and theory: the measured positions of each minimum and maximum are aligned with those of the theory, consistent with low coherent-error rates. In addition, we observe excellent agreement between the absolute values at the minima and maxima, indicating low incoherent-error rates as well. Even with high gate fidelities, a high algorithmic fidelity is not guaranteed. Randomized benchmarking gives the average fidelity over a large number of random gates, which transforms any coherent errors into incoherent ones. For real quantum algorithm circuits, the gates are generally not random. Therefore, any coherent errors can quickly add up and yield algorithmic performance far lower than expected from randomized benchmarking fidelities alone \cite{michielsen2017benchmarking,kjaergaard2020quantum}.

To quantify the performance of QAOA with $p=1$, we compare the highest-probability state at the minima of $F$ with the solutions in \cref{tab:problems}.
Problem A [\cref{fig:linecuts}(a)] does not reach its ground state ($F\approx -0.5$); however, the probability of measuring the correct state ($\ket{10}$) is approximately 50\%, which is still better than random guessing.
For problem C [\cref{fig:linecuts}(c)], we see that $F\approx -1$ does indeed correspond to a probability close to unity of measuring the ground state ($\ket{11}$). 
Problems B and D [\cref{fig:linecuts}(b) and (d)] have degenerate ground states, indicated by two state probabilities close to 50\% each at $F\approx -1$.

\section{Increasing the success probability}
To increase the success probability for problem A, we add an additional level ($p=2$). For $p>1$, a grid search to map out the full landscape becomes unfeasible due to the many parameters (equal to $2p$). Therefore, we instead use black-box optimizers to find the optimal variational angles. We try three different gradient-free optimizers: Bayesian optimization with Gaussian processes (BGP), Nelder-Mead, and covariance matrix adaptation evolution strategy (CMA-ES). We choose BGP due to its ability to find global minima, Nelder-Mead due to it being common and simple, and CMA-ES due to its favorable scaling with the number of optimization parameters.

We evaluate the optimizer performances by running 200 independent optimizations with random starting values ($\vec{\gamma},\vec{\beta}\in[0,\pi[$) for each optimizer. For each set of variational angles, we repeat the circuit and measure 5000 samples to accurately estimate the expectation values.
We set a threshold for convergence at $F<-0.95$ and count the number of converged optimization runs as well as the number of calls to the quantum processor (function calls) required to converge. We also record the success probability of measuring the problem solution ($P_{\ket{10}}$). The results are summarized in \cref{tab:opt}. 

We observe that the success probabilities after convergence are similar for all three optimizers. However, there is a difference in convergence probability, of which BGP has the highest and Nelder-Mead has the lowest. The lower performance of Nelder-Mead is most likely due to its sensitivity to local minima, a well-known problem for most optimizers. In contrast, one of the strengths of Bayesian optimization is its ability to find global minima, which could explain why it performs better than Nelder-Mead and CMA-ES. Additionally, Bayesian optimization is designed to handle optimization where the time of each function call is high (costly), such that the number of call is kept low. However, for more optimization parameters (higher $p$), the performance of BGP is generally decreased due to an increasing need for classical computation. CMA-ES, on the other hand, excels when the number of parameters are high, and thus might be a good optimizer for QAOA with tens or hundreds of parameters. Here, with just four parameters, CMA-ES has a convergence-probability similar to that of BGP, although with a greater number of function calls on average. 

\begin{table}
\centering
\caption{Comparison between different optimizers. We run QAOA for problem A over 200 iterations with random starting parameters. We extract the convergence probability for reaching a cost below -0.95, the average number of function calls required to reach that level, and the highest achieved probability of measuring the problem solution ($P_{\ket{10}}$).
}
\label{tab:opt}
\begin{ruledtabular}
\begin{tabular}{ p{2cm} | p{2cm} p{2cm} p{2cm}}
  Optimizer & Convergence & Function calls & $P_{\ket{10}}$ \\
  \hline
  BGP & 61.5\% & $44 \pm 16$ & 96.5\% \\
  Nelder-Mead & 20.0\% & $38 \pm 13$ & 96.3\% \\
  CMA-ES & 49.5\% & $121 \pm 46$ & 96.6\% \\
\end{tabular}
\end{ruledtabular}
\end{table}

\begin{figure}
    \centering
    \includegraphics[width=1\columnwidth]{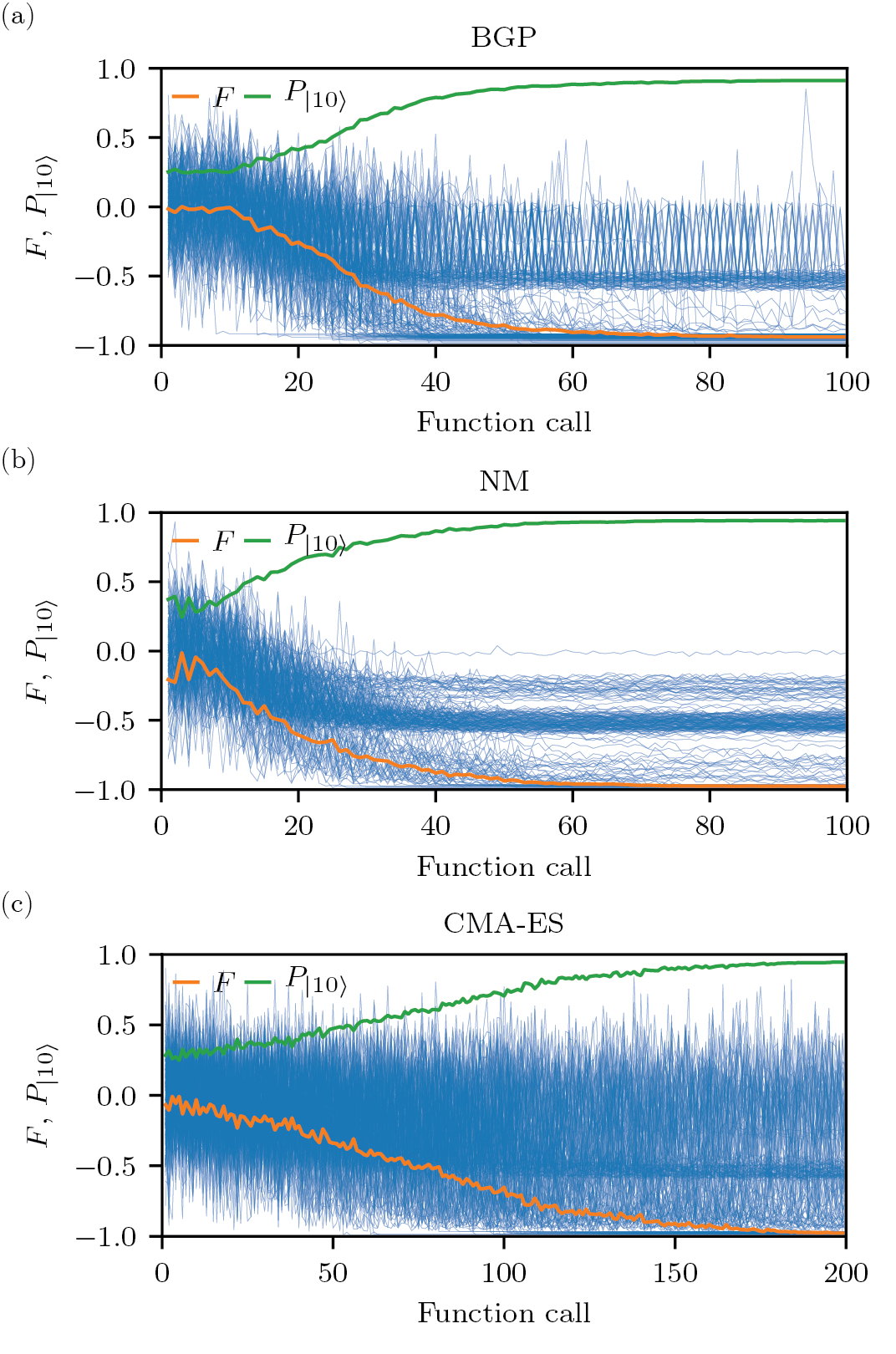}
    \caption{Optimization of variational angles for problem A using $p=2$ iterations of the algorithm and three different black-box and gradient-free optimizers:
    (a) Bayesian optimization with Gaussian processes, (b) Nelder-Mead, and (c) covariance matrix adaptation evolution strategy.
    We run the optimization 200 times with random starting parameters. Plotted as blue lines are the individual optimization trajectories for $F$, where each data-point is the average of 5000 measurements. In orange and green are the cost ($F)$ and success probability ($P_{\ket{10}}$) averaged over the converged runs.}
    \label{fig:opt}
\end{figure}

To quantify the optimization further, we study the trajectories of each optimization run (\cref{fig:opt}).  For each run, we plot the costs $F$. The trajectories for BGP and Nelder-Mead [\cref{fig:opt}(a-b)] corroborate the indications about local minima. We see groups of horizontal lines corresponding to different local minima, especially clear at $F\approx -0.55$ for both BGP and Nelder-Mead. We also see that BGP tries, and sometimes succeeds, to escape these local minima, which is one of the advantages of Bayesian optimization. In comparison, Nelder-Mead rarely gets out of a local minimum once it has found it. For the third optimizer, CMA-ES [\cref{fig:opt}(c)], it is hard to draw any conclusions from the trajectories other than that the convergence is slower than for the other optimizers. However, we include the CMA-ES trajectories for completeness. For each optimizer, we also plot the averaged (over all the converged) trajectories for $F$ and the probability of finding the solution state $P_{\ket{10}}$.

\begin{figure}
    \centering
    \includegraphics[width=1\columnwidth]{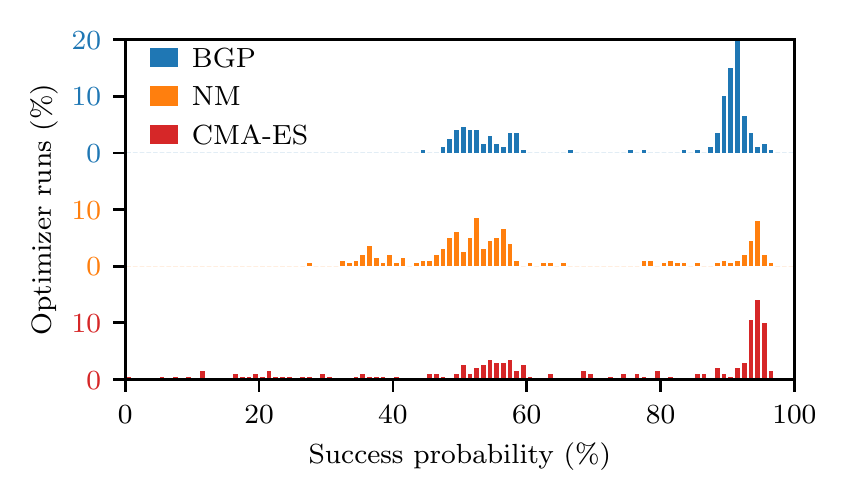}
    \caption{Histogram of the final success probabilities ($P_{\ket{10}}$) for three different optimizers on problem A using $p=2$ iterations of the algorithm. Each optimizer was run 200 times. The bars are vertically offset for clarity.
    }
    \label{fig:hist}
\end{figure}

At the end of the optimization, the highest recorded probability of generating the correct state is 96.6\%. The success probability is limited by imperfect gates (we have verified that an ideal quantum computer and $p=2$ can achieve $P_{\ket{10}}=1$).  We compare our measured success probability to what we would expect from the randomized-benchmarking fidelities. The quantum circuit for $p=2$ consists of 6 X, 4 Hadamard, 4 Z, and 3 CZ gates, which, when multiplied together with the fidelities for each gate, predicts a total fidelity of 96.3\%, in good agreement with the measured fidelity considering experimental uncertainties (e.g., fluctuations in qubit coherence and gate fidelities). Note that $p=3$ would not yield a higher success probability, since adding more gates would lower the total fidelity further (predicted to be 94.2\%).  

Finally, we examine histograms over the success probabilities at the end of each optimization run for the three different optimizers, see \cref{fig:hist}. Again, we observe that BGP has the most converged runs out of the three. We see clusters around 55 and \SI{95}{\percent} success probability for all three optimizers, possibly corresponding to one local and the global minima. For CMA-ES the success probabilities are more scattered, where some runs even have below \SI{40}{\percent} success. All in all, Bayesian optimization performs the best; however, further studies will be needed on which classical optimizer is the most suitable for variational quantum algorithms, such as QAOA and VQE.

\section{Conclusion}
In conclusion, we have implemented the quantum approximate optimization algorithm with up to $p=2$ levels. Using a superconducting quantum processor with state-of-the-art performance, we successfully optimized four instances of the exact-cover problem. For the non-trivial instance (problem A), we used $p=2$ and black-box optimization to reach a success probability to 96.6\% (up from 50\% with $p=1$), in good agreement with a prediction from our gate fidelities. 
Even if many more qubits are needed to solve problems that are intractable for classical computers, algorithmic performance serves as a critical quantum-processor benchmark since performance can be much lower than what individual gate fidelities predicts. 
Although further experiments with larger devices are needed to explore if QAOA can have an advantage over classical algorithms, our results show that QAOA can be used to solve the exact-cover problem. 

\begin{acknowledgments}
We are grateful to the Quantum Device Lab at ETH Zürich for sharing their designs of sample holder and printed circuit board, and Simon Gustavsson and Bruno Küng for valuable support with the measurement infrastructure. We also thank Morten Kjaergaard, Devdatt Dubhashi, and Kevin Pack for insightful discussions. This work was performed in part at Myfab Chalmers. We acknowledge financial support from the Knut and Alice Wallenberg Foundation, the Swedish Research Council, and the EU Flagship on Quantum Technology H2020-FETFLAG-2018-03 project 820363 OpenSuperQ.
\end{acknowledgments}

\begin{appendix}
\section{Measurement setup}
\label{app:setup}
\begin{figure}
    \centering
    \includegraphics[width=1\linewidth]{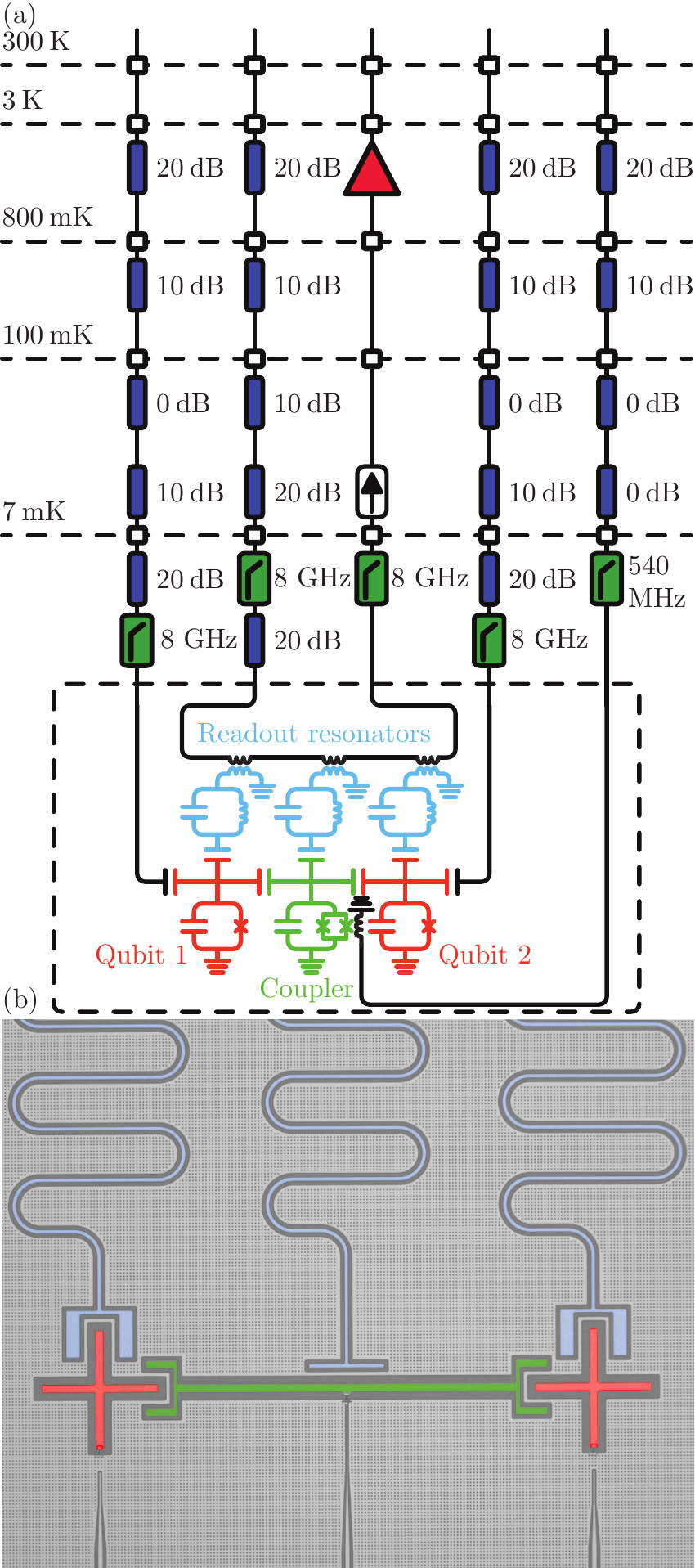}
    \caption{(a) Cryogenic setup and electrical circuit of the quantum processor. All lines are attenuated and filtered to minimize the amount of noise reaching the qubits. The readout output contains cryogenic isolators and a high electron mobility transistor amplifier. 
    (b) False-colored micro-graph of the processor. The colors match the circuit elements in (a). The three waveguides at the bottom are for control over the qubits and the coupler.}
    \label{fig:setup}
\end{figure}

The experimental measurement setup used here is a standard circuit-QED setup, see schematic in \cref{fig:setup}. The quantum processor consists of two xmon-style transmon qubits coupled via a frequency-tunable anharmonic oscillator. The tunability is provided by two Josephson junctions in a SQUID configuration. The two qubits are capacitively coupled to individual control lines and quarter-wavelength resonators for readout. There is also a readout resonator for the coupler, which is only used as a debugging tool (i.e., it is not involved during any algorithm execution).
The SQUID for the tunable coupler is inductively coupled to a waveguide to allow for both static and fast modulation of the resonant frequency.

The processor is fabricated on a high-resistivity intrinsic silicon substrate. After initial chemical cleaning, an aluminium film is evaporated. All features except the Josephson junctions are patterned by direct-write laser lithography and etched with a warm mixture of acids. The Josephson junctions are patterned by electron-beam lithography, and evaporated from the same target as previously. A third lithography and evaporation step (with in-situ ion milling) is performed to connect the Josephson junctions to the rest of the circuit. Finally, the wafer is diced into individual dies and subsequently cleaned by a combination of wet and dry chemistry.

Then, a die is selected and packaged in a copper box and wire bonded to a palladium- and gold-plated printed circuit board with 16 non-magnetic coaxial connectors. For the present device, we use 5 of these connectors, 2 for readout, 2 for single-qubit control, and 1 for control of the magnetic flux through the SQUID loop of the coupler. These are connected to filtered and attenuated coaxial lines leading up to room temperature. We point out that the DC current for the static flux bias is also provided through the coaxial line. Finally, the processor is attached to the mixing chamber of a Bluefors LD250 cryo-free dilution refrigerator. There, it is shielded from stray magnetic fields by two shields of cryoperm/mu-metal and two shields of superconductors.

We perform multiplexed readout by using a Zurich Instruments UHFQA for generating and detecting the readout signals, together with a Rohde \& Schwarz SGS100A continuous-wave signal generator and two Marki IQ mixers for up- and down-conversion. The single-qubit pulses are synthesized by a Zurich Instruments HDAWG and upconverted using Rohde \& Schwarz SGS100A vector signal generators. The flux drive is generated directly by the HDAWG since the modulation frequency is within the bandwidth of the instrument. Finally, all instruments are controlled and orchestrated by the measurement and automation software Labber. Labber also does cost-function evaluations and calls external Python packages for the three different optimizers. All three optimizers were run using publicly available packages: Scikit-Optimize for BGP, scipy for Nelder-Mead, and pycma for CMA-ES.

\section{Characterization and tune-up}
\label{app:char}
Initially, we perform basic spectroscopy and decoherence benchmarking of each qubit individually. This allows us to extract readout frequencies, qubit frequencies and anharmonicities, relaxation and dephasing times, and static couplings between qubit and resonator, as well as between qubit and coupler. The extracted parameters are found in \cref{tab:device}.

\begin{table}
\centering
\caption{Device parameters. Readout-resonator frequency $f_R$ and qubit transition frequencies $f_{ij}$. $g$ is the coupling between qubit and resonator, and $j$ is the coupling between qubit and coupler. $T_1$ and $T_2^*$ are the relaxation and free induction decay times measured over 14  hours. $F_{1q}$, $F_m$, and $F_{CZ}$ are the single-qubit, measurement, and CZ fidelities, respectively.
}
\label{tab:device}
\begin{ruledtabular}
\begin{tabular}{ p{1.6cm} | p{1.4cm} p{1cm} p{1.4cm}}
  Parameter & Qubit 1 & & Qubit 2 \\
  \hline
  $f_R$ & 6.17~GHz & & 6.04~GHz \\
  $f_{01}$ & 3.82~GHz & & 4.30~GHz \\
  $f_{12}-f_{01}$ & -229~MHz & & -225~MHz \\
  $j$ & 29.1~MHz & & 33.0~MHz \\
  $g$ & 53.2~MHz & & 56.9~MHz \\
  $T_1$ & $77$~µs & & $55$~µs \\
  $T_2^*$ & 49~µs & & 82~µs  \\
  $F_{1q}$ & 0.9986 & & 0.9993  \\
  $F_m$ & 0.86 & & 0.95 \\
  $F_{CZ}$ & & 0.986 &   \\
\end{tabular}
\end{ruledtabular}
\end{table}

After the initial characterization, we tune up high-fidelity single-qubit gates. The drive pulses have cosine envelopes together with first-order DRAG components to compensate for the qubit frequency shift due to the driving. Our rather long (\SI{50}{\nano\second}) pulses makes leakage from $\ket{1}$ to $\ket{2}$ minimal even without DRAG. To find optimal pulse amplitudes and DRAG coefficients, we use error amplification by applying varying lengths of trains of $\pi$ pulses. Qubit drive frequencies are measured accurately by detuned Ramsey fringes.

Next, we calibrate our readout fidelities. By collecting raw voltages of the readout signals (as measured by the digitizer in the UHFQA), with and without a calibrated pi-pulse applied to the qubit ($|0\rangle$ and $|1\rangle$ states, respectively), and as a function of readout frequency and amplitude, we can find the optimal readout parameters. Due to our rather low coupling strengths, we cannot achieve short readout times in this device. However, QAOA does not require any measurement feedback, so a long readout time is not an issue as long as the time is shorter than the relaxation times of the qubits. Also, longer readout times give greater signal-to-noise ratios, which allows us to achieve high readout fidelities even in the absence of a quantum-limited amplifier. Here, the readout is \SI{2.3}{\micro\second} long, well below our relaxation times (several tens of microseconds). After finding the optimal readout parameters, a voltage threshold is used to differentiate between $\ket{0}$ or $\ket{1}$ of the measured qubit. 

To accurately extract state probabilities in the presence of limited readout fidelity, we collect statistics of the measured qubit population as a function of qubit drive amplitude (Rabi oscillations). Since the measured population increases monotonically with the expected population, we can renormalize the populations, similarly to Ref.~\cite{chow2010detecting}. This calibration allows us to accurately measure the average quantities $\langle \sigma_i^z \rangle$, $\langle \sigma_i^1\sigma_i^2  \rangle$ and state probabilities even in the presence of limited readout fidelities.

Our two-qubit gate of choice is the controlled phase (CZ). This interaction is induced by parametrically modulating the resonant frequency of the coupler at a frequency close to the difference of $\ket{11}$ and $\ket{02}$. For our device, this frequency is \SI{255}{\mega\hertz}. However, due to the frequency modulation and the non-linear relationship between flux and frequency, the transition frequencies are slightly lowered. This frequency shift will also induce deterministic phase shifts on the individual qubits, which we compensate for by applying Z gates on both qubits after each CZ gate. We choose a static bias point and a modulation amplitude that yield a moderate effective coupling strength of \SI{5}{\mega\hertz} between the two states. From here, we find the modulation frequency and time that yield a full oscillation between the $\ket{11}$ and $\ket{02}$ states. We then fine-tune the frequency and time such that the controlled phase is $\pi$ and the leakage to $\ket{02}$ is minimal. Here, the final gate frequency and duration were \SI{253}{\mega\hertz} and \SI{271}{\nano\second}.

\begin{figure}
    \centering
    \includegraphics[width=1\linewidth]{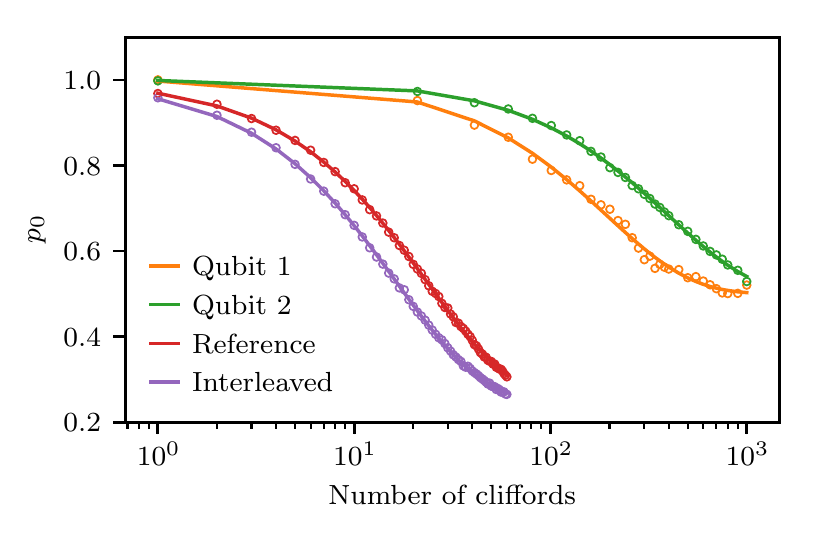}
    \caption{Randomized benchmarking of single- and two-qubit gates. Plotted are the probability of measuring the ground state as a function of number of Clifford gates applied. Circles are data, and lines are fits to extract the gate fidelity. For qubit 1 and 2, the extracted single-qubit fidelities (averaged over all possible single-qubit Cliffords) are 0.9986 and 0.9993. For benchmarking of the two-qubit gate, we take a reference (random Clifford gates) and an interleaved (a CZ gate between each Clifford) trace to extract the CZ fidelity (0.986).}
    \label{fig:gates}
\end{figure}

We benchmark our single and two-qubit fidelities using randomized benchmarking. A sequence of random gates drawn from the Clifford group is applied together with a final recovery gate which should take the system back to the ground state. The number of random gates is varied and the probability of measuring the ground state is recorded. In \cref{fig:gates}, we plot these probabilities for each qubit individually and for the two-qubit case. In the single-qubit case, it is important to note that it was done simultaneously for both qubits. Generally, the gate fidelities are higher if they are done in isolation. However, to reduce the total run time of algorithms, we usually run single-qubit gates in parallel. Therefore, simultaneous randomized benchmarking fidelities are more relevant metrics than isolated ones. 

\end{appendix}
\bibliography{bib}

\begin{thebibliography}{31}%
\makeatletter
\providecommand \@ifxundefined [1]{%
 \@ifx{#1\undefined}
}%
\providecommand \@ifnum [1]{%
 \ifnum #1\expandafter \@firstoftwo
 \else \expandafter \@secondoftwo
 \fi
}%
\providecommand \@ifx [1]{%
 \ifx #1\expandafter \@firstoftwo
 \else \expandafter \@secondoftwo
 \fi
}%
\providecommand \natexlab [1]{#1}%
\providecommand \enquote  [1]{``#1''}%
\providecommand \bibnamefont  [1]{#1}%
\providecommand \bibfnamefont [1]{#1}%
\providecommand \citenamefont [1]{#1}%
\providecommand \href@noop [0]{\@secondoftwo}%
\providecommand \href [0]{\begingroup \@sanitize@url \@href}%
\providecommand \@href[1]{\@@startlink{#1}\@@href}%
\providecommand \@@href[1]{\endgroup#1\@@endlink}%
\providecommand \@sanitize@url [0]{\catcode `\\12\catcode `\$12\catcode
  `\&12\catcode `\#12\catcode `\^12\catcode `\_12\catcode `\%12\relax}%
\providecommand \@@startlink[1]{}%
\providecommand \@@endlink[0]{}%
\providecommand \url  [0]{\begingroup\@sanitize@url \@url }%
\providecommand \@url [1]{\endgroup\@href {#1}{\urlprefix }}%
\providecommand \urlprefix  [0]{URL }%
\providecommand \Eprint [0]{\href }%
\providecommand \doibase [0]{https://doi.org/}%
\providecommand \selectlanguage [0]{\@gobble}%
\providecommand \bibinfo  [0]{\@secondoftwo}%
\providecommand \bibfield  [0]{\@secondoftwo}%
\providecommand \translation [1]{[#1]}%
\providecommand \BibitemOpen [0]{}%
\providecommand \bibitemStop [0]{}%
\providecommand \bibitemNoStop [0]{.\EOS\space}%
\providecommand \EOS [0]{\spacefactor3000\relax}%
\providecommand \BibitemShut  [1]{\csname bibitem#1\endcsname}%
\let\auto@bib@innerbib\@empty
\bibitem [{\citenamefont {Montanaro}(2016)}]{montanaro2016quantum}%
  \BibitemOpen
  \bibfield  {author} {\bibinfo {author} {\bibfnamefont {A.}~\bibnamefont
  {Montanaro}},\ }\bibfield  {title} {\bibinfo {title} {Quantum algorithms: an
  overview},\ }\href@noop {} {\bibfield  {journal} {\bibinfo  {journal} {npj
  Quantum Information}\ }\textbf {\bibinfo {volume} {2}},\ \bibinfo {pages}
  {15023} (\bibinfo {year} {2016})}\BibitemShut {NoStop}%
\bibitem [{\citenamefont {Wendin}(2017)}]{wendin2017quantum}%
  \BibitemOpen
  \bibfield  {author} {\bibinfo {author} {\bibfnamefont {G.}~\bibnamefont
  {Wendin}},\ }\bibfield  {title} {\bibinfo {title} {Quantum information
  processing with superconducting circuits: a review},\ }\href@noop {}
  {\bibfield  {journal} {\bibinfo  {journal} {Reports on Progress in Physics}\
  }\textbf {\bibinfo {volume} {80}},\ \bibinfo {pages} {106001} (\bibinfo
  {year} {2017})}\BibitemShut {NoStop}%
\bibitem [{\citenamefont {Gu}\ \emph {et~al.}(2017)\citenamefont {Gu},
  \citenamefont {Kockum}, \citenamefont {Miranowicz}, \citenamefont {Liu},\
  and\ \citenamefont {Nori}}]{GU20171}%
  \BibitemOpen
  \bibfield  {author} {\bibinfo {author} {\bibfnamefont {X.}~\bibnamefont
  {Gu}}, \bibinfo {author} {\bibfnamefont {A.~F.}\ \bibnamefont {Kockum}},
  \bibinfo {author} {\bibfnamefont {A.}~\bibnamefont {Miranowicz}}, \bibinfo
  {author} {\bibfnamefont {Y.~X.}\ \bibnamefont {Liu}},\ and\ \bibinfo {author}
  {\bibfnamefont {F.}~\bibnamefont {Nori}},\ }\bibfield  {title} {\bibinfo
  {title} {Microwave photonics with superconducting quantum circuits},\ }\href
  {https://doi.org/https://doi.org/10.1016/j.physrep.2017.10.002} {\bibfield
  {journal} {\bibinfo  {journal} {Physics Reports}\ }\textbf {\bibinfo {volume}
  {718-719}},\ \bibinfo {pages} {1 } (\bibinfo {year} {2017})}\BibitemShut
  {NoStop}%
\bibitem [{\citenamefont {Kjaergaard}\ \emph
  {et~al.}(2020{\natexlab{a}})\citenamefont {Kjaergaard}, \citenamefont
  {Schwartz}, \citenamefont {Braum{\"u}ller}, \citenamefont {Krantz},
  \citenamefont {Wang}, \citenamefont {Gustavsson},\ and\ \citenamefont
  {Oliver}}]{kjaergaard2019superconducting}%
  \BibitemOpen
  \bibfield  {author} {\bibinfo {author} {\bibfnamefont {M.}~\bibnamefont
  {Kjaergaard}}, \bibinfo {author} {\bibfnamefont {M.~E.}\ \bibnamefont
  {Schwartz}}, \bibinfo {author} {\bibfnamefont {J.}~\bibnamefont
  {Braum{\"u}ller}}, \bibinfo {author} {\bibfnamefont {P.}~\bibnamefont
  {Krantz}}, \bibinfo {author} {\bibfnamefont {J.~I.-J.}\ \bibnamefont {Wang}},
  \bibinfo {author} {\bibfnamefont {S.}~\bibnamefont {Gustavsson}},\ and\
  \bibinfo {author} {\bibfnamefont {W.~D.}\ \bibnamefont {Oliver}},\ }\bibfield
   {title} {\bibinfo {title} {Superconducting qubits: Current state of play},\
  }\href@noop {} {\bibfield  {journal} {\bibinfo  {journal} {Annual Review of
  Condensed Matter Physics}\ }\textbf {\bibinfo {volume} {11}},\ \bibinfo
  {pages} {369} (\bibinfo {year} {2020}{\natexlab{a}})}\BibitemShut {NoStop}%
\bibitem [{\citenamefont {Preskill}(2018)}]{Perskillnisq}%
  \BibitemOpen
  \bibfield  {author} {\bibinfo {author} {\bibfnamefont {J.}~\bibnamefont
  {Preskill}},\ }\bibfield  {title} {\bibinfo {title} {Quantum computing in the
  nisq era and beyond},\ }\href@noop {} {\bibfield  {journal} {\bibinfo
  {journal} {Quantum}\ }\textbf {\bibinfo {volume} {2}},\ \bibinfo {pages} {79}
  (\bibinfo {year} {2018})}\BibitemShut {NoStop}%
\bibitem [{\citenamefont {Boixo}\ \emph {et~al.}(2018)\citenamefont {Boixo},
  \citenamefont {Isakov}, \citenamefont {Smelyanskiy}, \citenamefont {Babbush},
  \citenamefont {Ding}, \citenamefont {Jiang}, \citenamefont {Bremner},
  \citenamefont {Martinis},\ and\ \citenamefont
  {Neven}}]{boixo2018characterizing}%
  \BibitemOpen
  \bibfield  {author} {\bibinfo {author} {\bibfnamefont {S.}~\bibnamefont
  {Boixo}}, \bibinfo {author} {\bibfnamefont {S.~V.}\ \bibnamefont {Isakov}},
  \bibinfo {author} {\bibfnamefont {V.~N.}\ \bibnamefont {Smelyanskiy}},
  \bibinfo {author} {\bibfnamefont {R.}~\bibnamefont {Babbush}}, \bibinfo
  {author} {\bibfnamefont {N.}~\bibnamefont {Ding}}, \bibinfo {author}
  {\bibfnamefont {Z.}~\bibnamefont {Jiang}}, \bibinfo {author} {\bibfnamefont
  {M.~J.}\ \bibnamefont {Bremner}}, \bibinfo {author} {\bibfnamefont {J.~M.}\
  \bibnamefont {Martinis}},\ and\ \bibinfo {author} {\bibfnamefont
  {H.}~\bibnamefont {Neven}},\ }\bibfield  {title} {\bibinfo {title}
  {Characterizing quantum supremacy in near-term devices},\ }\href@noop {}
  {\bibfield  {journal} {\bibinfo  {journal} {Nature Physics}\ }\textbf
  {\bibinfo {volume} {14}},\ \bibinfo {pages} {595} (\bibinfo {year}
  {2018})}\BibitemShut {NoStop}%
\bibitem [{\citenamefont {Arute}\ \emph {et~al.}(2019)\citenamefont {Arute},
  \citenamefont {Arya}, \citenamefont {Babbush}, \citenamefont {Bacon},
  \citenamefont {Bardin}, \citenamefont {Barends}, \citenamefont {Biswas},
  \citenamefont {Boixo}, \citenamefont {Brandao}, \citenamefont {Buell} \emph
  {et~al.}}]{arute2019quantum}%
  \BibitemOpen
  \bibfield  {author} {\bibinfo {author} {\bibfnamefont {F.}~\bibnamefont
  {Arute}}, \bibinfo {author} {\bibfnamefont {K.}~\bibnamefont {Arya}},
  \bibinfo {author} {\bibfnamefont {R.}~\bibnamefont {Babbush}}, \bibinfo
  {author} {\bibfnamefont {D.}~\bibnamefont {Bacon}}, \bibinfo {author}
  {\bibfnamefont {J.~C.}\ \bibnamefont {Bardin}}, \bibinfo {author}
  {\bibfnamefont {R.}~\bibnamefont {Barends}}, \bibinfo {author} {\bibfnamefont
  {R.}~\bibnamefont {Biswas}}, \bibinfo {author} {\bibfnamefont
  {S.}~\bibnamefont {Boixo}}, \bibinfo {author} {\bibfnamefont {F.~G.}\
  \bibnamefont {Brandao}}, \bibinfo {author} {\bibfnamefont {D.~A.}\
  \bibnamefont {Buell}}, \emph {et~al.},\ }\bibfield  {title} {\bibinfo {title}
  {Quantum supremacy using a programmable superconducting processor},\
  }\href@noop {} {\bibfield  {journal} {\bibinfo  {journal} {Nature}\ }\textbf
  {\bibinfo {volume} {574}},\ \bibinfo {pages} {505} (\bibinfo {year}
  {2019})}\BibitemShut {NoStop}%
\bibitem [{\citenamefont {Farhi}\ \emph {et~al.}(2014)\citenamefont {Farhi},
  \citenamefont {Goldstone},\ and\ \citenamefont {Gutmann}}]{farhi2014quantum}%
  \BibitemOpen
  \bibfield  {author} {\bibinfo {author} {\bibfnamefont {E.}~\bibnamefont
  {Farhi}}, \bibinfo {author} {\bibfnamefont {J.}~\bibnamefont {Goldstone}},\
  and\ \bibinfo {author} {\bibfnamefont {S.}~\bibnamefont {Gutmann}},\
  }\bibfield  {title} {\bibinfo {title} {A quantum approximate optimization
  algorithm},\ }\href@noop {} {\bibfield  {journal} {\bibinfo  {journal} {arXiv
  preprint arXiv:1411.4028}\ } (\bibinfo {year} {2014})}\BibitemShut {NoStop}%
\bibitem [{\citenamefont {Otterbach}\ \emph {et~al.}(2017)\citenamefont
  {Otterbach}, \citenamefont {Manenti}, \citenamefont {Alidoust}, \citenamefont
  {Bestwick}, \citenamefont {Block}, \citenamefont {Bloom}, \citenamefont
  {Caldwell}, \citenamefont {Didier}, \citenamefont {Fried}, \citenamefont
  {Hong} \emph {et~al.}}]{otterbach2017unsupervised}%
  \BibitemOpen
  \bibfield  {author} {\bibinfo {author} {\bibfnamefont {J.}~\bibnamefont
  {Otterbach}}, \bibinfo {author} {\bibfnamefont {R.}~\bibnamefont {Manenti}},
  \bibinfo {author} {\bibfnamefont {N.}~\bibnamefont {Alidoust}}, \bibinfo
  {author} {\bibfnamefont {A.}~\bibnamefont {Bestwick}}, \bibinfo {author}
  {\bibfnamefont {M.}~\bibnamefont {Block}}, \bibinfo {author} {\bibfnamefont
  {B.}~\bibnamefont {Bloom}}, \bibinfo {author} {\bibfnamefont
  {S.}~\bibnamefont {Caldwell}}, \bibinfo {author} {\bibfnamefont
  {N.}~\bibnamefont {Didier}}, \bibinfo {author} {\bibfnamefont {E.~S.}\
  \bibnamefont {Fried}}, \bibinfo {author} {\bibfnamefont {S.}~\bibnamefont
  {Hong}}, \emph {et~al.},\ }\bibfield  {title} {\bibinfo {title} {Unsupervised
  machine learning on a hybrid quantum computer},\ }\href@noop {} {\bibfield
  {journal} {\bibinfo  {journal} {arXiv preprint arXiv:1712.05771}\ } (\bibinfo
  {year} {2017})}\BibitemShut {NoStop}%
\bibitem [{\citenamefont {Pagano}\ \emph {et~al.}(2019)\citenamefont {Pagano},
  \citenamefont {Bapat}, \citenamefont {Becker}, \citenamefont {Collins},
  \citenamefont {De}, \citenamefont {Hess}, \citenamefont {Kaplan},
  \citenamefont {Kyprianidis}, \citenamefont {Tan}, \citenamefont {Baldwin}
  \emph {et~al.}}]{pagano2019quantum}%
  \BibitemOpen
  \bibfield  {author} {\bibinfo {author} {\bibfnamefont {G.}~\bibnamefont
  {Pagano}}, \bibinfo {author} {\bibfnamefont {A.}~\bibnamefont {Bapat}},
  \bibinfo {author} {\bibfnamefont {P.}~\bibnamefont {Becker}}, \bibinfo
  {author} {\bibfnamefont {K.}~\bibnamefont {Collins}}, \bibinfo {author}
  {\bibfnamefont {A.}~\bibnamefont {De}}, \bibinfo {author} {\bibfnamefont
  {P.}~\bibnamefont {Hess}}, \bibinfo {author} {\bibfnamefont {H.}~\bibnamefont
  {Kaplan}}, \bibinfo {author} {\bibfnamefont {A.}~\bibnamefont {Kyprianidis}},
  \bibinfo {author} {\bibfnamefont {W.}~\bibnamefont {Tan}}, \bibinfo {author}
  {\bibfnamefont {C.}~\bibnamefont {Baldwin}}, \emph {et~al.},\ }\bibfield
  {title} {\bibinfo {title} {Quantum approximate optimization with a
  trapped-ion quantum simulator},\ }\href@noop {} {\bibfield  {journal}
  {\bibinfo  {journal} {arXiv preprint arXiv:1906.02700}\ } (\bibinfo {year}
  {2019})}\BibitemShut {NoStop}%
\bibitem [{\citenamefont {Peruzzo}\ \emph {et~al.}(2014)\citenamefont
  {Peruzzo}, \citenamefont {McClean}, \citenamefont {Shadbolt}, \citenamefont
  {Yung}, \citenamefont {Zhou}, \citenamefont {Love}, \citenamefont
  {Aspuru-Guzik},\ and\ \citenamefont {O’Brien}}]{peruzzo2014variational}%
  \BibitemOpen
  \bibfield  {author} {\bibinfo {author} {\bibfnamefont {A.}~\bibnamefont
  {Peruzzo}}, \bibinfo {author} {\bibfnamefont {J.}~\bibnamefont {McClean}},
  \bibinfo {author} {\bibfnamefont {P.}~\bibnamefont {Shadbolt}}, \bibinfo
  {author} {\bibfnamefont {M.-H.}\ \bibnamefont {Yung}}, \bibinfo {author}
  {\bibfnamefont {X.-Q.}\ \bibnamefont {Zhou}}, \bibinfo {author}
  {\bibfnamefont {P.~J.}\ \bibnamefont {Love}}, \bibinfo {author}
  {\bibfnamefont {A.}~\bibnamefont {Aspuru-Guzik}},\ and\ \bibinfo {author}
  {\bibfnamefont {J.~L.}\ \bibnamefont {O’Brien}},\ }\bibfield  {title}
  {\bibinfo {title} {A variational eigenvalue solver on a photonic quantum
  processor},\ }\href@noop {} {\bibfield  {journal} {\bibinfo  {journal}
  {Nature Communications}\ }\textbf {\bibinfo {volume} {5}},\ \bibinfo {pages}
  {4213} (\bibinfo {year} {2014})}\BibitemShut {NoStop}%
\bibitem [{\citenamefont {O’Malley}\ \emph {et~al.}(2016)\citenamefont
  {O’Malley}, \citenamefont {Babbush}, \citenamefont {Kivlichan},
  \citenamefont {Romero}, \citenamefont {McClean}, \citenamefont {Barends},
  \citenamefont {Kelly}, \citenamefont {Roushan}, \citenamefont {Tranter},
  \citenamefont {Ding} \emph {et~al.}}]{o2016scalable}%
  \BibitemOpen
  \bibfield  {author} {\bibinfo {author} {\bibfnamefont {P.~J.~J.}\
  \bibnamefont {O’Malley}}, \bibinfo {author} {\bibfnamefont
  {R.}~\bibnamefont {Babbush}}, \bibinfo {author} {\bibfnamefont {I.~D.}\
  \bibnamefont {Kivlichan}}, \bibinfo {author} {\bibfnamefont {J.}~\bibnamefont
  {Romero}}, \bibinfo {author} {\bibfnamefont {J.~R.}\ \bibnamefont {McClean}},
  \bibinfo {author} {\bibfnamefont {R.}~\bibnamefont {Barends}}, \bibinfo
  {author} {\bibfnamefont {J.}~\bibnamefont {Kelly}}, \bibinfo {author}
  {\bibfnamefont {P.}~\bibnamefont {Roushan}}, \bibinfo {author} {\bibfnamefont
  {A.}~\bibnamefont {Tranter}}, \bibinfo {author} {\bibfnamefont
  {N.}~\bibnamefont {Ding}}, \emph {et~al.},\ }\bibfield  {title} {\bibinfo
  {title} {Scalable quantum simulation of molecular energies},\ }\href@noop {}
  {\bibfield  {journal} {\bibinfo  {journal} {Physical Review X}\ }\textbf
  {\bibinfo {volume} {6}},\ \bibinfo {pages} {031007} (\bibinfo {year}
  {2016})}\BibitemShut {NoStop}%
\bibitem [{\citenamefont {Kandala}\ \emph {et~al.}(2017)\citenamefont
  {Kandala}, \citenamefont {Mezzacapo}, \citenamefont {Temme}, \citenamefont
  {Takita}, \citenamefont {Brink}, \citenamefont {Chow},\ and\ \citenamefont
  {Gambetta}}]{kandala2017hardware}%
  \BibitemOpen
  \bibfield  {author} {\bibinfo {author} {\bibfnamefont {A.}~\bibnamefont
  {Kandala}}, \bibinfo {author} {\bibfnamefont {A.}~\bibnamefont {Mezzacapo}},
  \bibinfo {author} {\bibfnamefont {K.}~\bibnamefont {Temme}}, \bibinfo
  {author} {\bibfnamefont {M.}~\bibnamefont {Takita}}, \bibinfo {author}
  {\bibfnamefont {M.}~\bibnamefont {Brink}}, \bibinfo {author} {\bibfnamefont
  {J.~M.}\ \bibnamefont {Chow}},\ and\ \bibinfo {author} {\bibfnamefont
  {J.~M.}\ \bibnamefont {Gambetta}},\ }\bibfield  {title} {\bibinfo {title}
  {Hardware-efficient variational quantum eigensolver for small molecules and
  quantum magnets},\ }\href@noop {} {\bibfield  {journal} {\bibinfo  {journal}
  {Nature}\ }\textbf {\bibinfo {volume} {549}},\ \bibinfo {pages} {242}
  (\bibinfo {year} {2017})}\BibitemShut {NoStop}%
\bibitem [{\citenamefont {Jiang}\ \emph {et~al.}(2017)\citenamefont {Jiang},
  \citenamefont {Rieffel},\ and\ \citenamefont {Wang}}]{jiang2017near}%
  \BibitemOpen
  \bibfield  {author} {\bibinfo {author} {\bibfnamefont {Z.}~\bibnamefont
  {Jiang}}, \bibinfo {author} {\bibfnamefont {E.~G.}\ \bibnamefont {Rieffel}},\
  and\ \bibinfo {author} {\bibfnamefont {Z.}~\bibnamefont {Wang}},\ }\bibfield
  {title} {\bibinfo {title} {Near-optimal quantum circuit for grover's
  unstructured search using a transverse field},\ }\href@noop {} {\bibfield
  {journal} {\bibinfo  {journal} {Physical Review A}\ }\textbf {\bibinfo
  {volume} {95}},\ \bibinfo {pages} {062317} (\bibinfo {year}
  {2017})}\BibitemShut {NoStop}%
\bibitem [{\citenamefont {Niu}\ \emph {et~al.}(2019)\citenamefont {Niu},
  \citenamefont {Lu},\ and\ \citenamefont {Chuang}}]{niu2019optimizing}%
  \BibitemOpen
  \bibfield  {author} {\bibinfo {author} {\bibfnamefont {M.~Y.}\ \bibnamefont
  {Niu}}, \bibinfo {author} {\bibfnamefont {S.}~\bibnamefont {Lu}},\ and\
  \bibinfo {author} {\bibfnamefont {I.~L.}\ \bibnamefont {Chuang}},\ }\bibfield
   {title} {\bibinfo {title} {Optimizing qaoa: Success probability and runtime
  dependence on circuit depth},\ }\href@noop {} {\bibfield  {journal} {\bibinfo
   {journal} {arXiv preprint arXiv:1905.12134}\ } (\bibinfo {year}
  {2019})}\BibitemShut {NoStop}%
\bibitem [{\citenamefont {Farhi}\ and\ \citenamefont
  {Harrow}(2016)}]{farhi2016quantum}%
  \BibitemOpen
  \bibfield  {author} {\bibinfo {author} {\bibfnamefont {E.}~\bibnamefont
  {Farhi}}\ and\ \bibinfo {author} {\bibfnamefont {A.~W.}\ \bibnamefont
  {Harrow}},\ }\bibfield  {title} {\bibinfo {title} {Quantum supremacy through
  the quantum approximate optimization algorithm},\ }\href@noop {} {\bibfield
  {journal} {\bibinfo  {journal} {arXiv preprint arXiv:1602.07674}\ } (\bibinfo
  {year} {2016})}\BibitemShut {NoStop}%
\bibitem [{\citenamefont {Gr{\"{o}}nkvist}(2005)}]{Gronkvist}%
  \BibitemOpen
  \bibfield  {author} {\bibinfo {author} {\bibfnamefont {M.}~\bibnamefont
  {Gr{\"{o}}nkvist}},\ }\emph {\bibinfo {title} {The Tail Assignment
  Problem}},\ \href@noop {} {Ph.D. thesis},\ \bibinfo  {school} {Chalmers
  University of Technology and G{\"{o}}teborg University} (\bibinfo {year}
  {2005})\BibitemShut {NoStop}%
\bibitem [{\citenamefont {Vikst{\aa}l}\ \emph {et~al.}(2019)\citenamefont
  {Vikst{\aa}l}, \citenamefont {Gr{\"o}nkvist}, \citenamefont {Svensson},
  \citenamefont {Andersson}, \citenamefont {Johansson},\ and\ \citenamefont
  {Ferrini}}]{Vikstal2019}%
  \BibitemOpen
  \bibfield  {author} {\bibinfo {author} {\bibfnamefont {P.}~\bibnamefont
  {Vikst{\aa}l}}, \bibinfo {author} {\bibfnamefont {M.}~\bibnamefont
  {Gr{\"o}nkvist}}, \bibinfo {author} {\bibfnamefont {M.}~\bibnamefont
  {Svensson}}, \bibinfo {author} {\bibfnamefont {M.}~\bibnamefont {Andersson}},
  \bibinfo {author} {\bibfnamefont {G.}~\bibnamefont {Johansson}},\ and\
  \bibinfo {author} {\bibfnamefont {G.}~\bibnamefont {Ferrini}},\ }\bibfield
  {title} {\bibinfo {title} {Applying the quantum approximate optimization
  algorithm to the tail assignment problem},\ }\href@noop {} {\bibfield
  {journal} {\bibinfo  {journal} {arXiv preprint arXiv:1912.10499}\ } (\bibinfo
  {year} {2019})}\BibitemShut {NoStop}%
\bibitem [{\citenamefont {Farhi}\ \emph {et~al.}(2001)\citenamefont {Farhi},
  \citenamefont {Goldstone}, \citenamefont {Gutmann}, \citenamefont {Lapan},
  \citenamefont {Lundgren},\ and\ \citenamefont {Preda}}]{farhi2001quantum}%
  \BibitemOpen
  \bibfield  {author} {\bibinfo {author} {\bibfnamefont {E.}~\bibnamefont
  {Farhi}}, \bibinfo {author} {\bibfnamefont {J.}~\bibnamefont {Goldstone}},
  \bibinfo {author} {\bibfnamefont {S.}~\bibnamefont {Gutmann}}, \bibinfo
  {author} {\bibfnamefont {J.}~\bibnamefont {Lapan}}, \bibinfo {author}
  {\bibfnamefont {A.}~\bibnamefont {Lundgren}},\ and\ \bibinfo {author}
  {\bibfnamefont {D.}~\bibnamefont {Preda}},\ }\bibfield  {title} {\bibinfo
  {title} {A quantum adiabatic evolution algorithm applied to random instances
  of an np-complete problem},\ }\href@noop {} {\bibfield  {journal} {\bibinfo
  {journal} {Science}\ }\textbf {\bibinfo {volume} {292}},\ \bibinfo {pages}
  {472} (\bibinfo {year} {2001})}\BibitemShut {NoStop}%
\bibitem [{\citenamefont {Wang}\ and\ \citenamefont
  {Wu}(2016)}]{wang2016ultrafast}%
  \BibitemOpen
  \bibfield  {author} {\bibinfo {author} {\bibfnamefont {H.}~\bibnamefont
  {Wang}}\ and\ \bibinfo {author} {\bibfnamefont {L.-A.}\ \bibnamefont {Wu}},\
  }\bibfield  {title} {\bibinfo {title} {Ultrafast adiabatic quantum algorithm
  for the np-complete exact cover problem},\ }\href@noop {} {\bibfield
  {journal} {\bibinfo  {journal} {Scientific reports}\ }\textbf {\bibinfo
  {volume} {6}},\ \bibinfo {pages} {22307} (\bibinfo {year}
  {2016})}\BibitemShut {NoStop}%
\bibitem [{\citenamefont {Gra\ss{}}(2019)}]{Grass2019quantum}%
  \BibitemOpen
  \bibfield  {author} {\bibinfo {author} {\bibfnamefont {T.}~\bibnamefont
  {Gra\ss{}}},\ }\bibfield  {title} {\bibinfo {title} {Quantum annealing with
  longitudinal bias fields},\ }\href
  {https://doi.org/10.1103/PhysRevLett.123.120501} {\bibfield  {journal}
  {\bibinfo  {journal} {Physical Review Letters}\ }\textbf {\bibinfo {volume}
  {123}},\ \bibinfo {pages} {120501} (\bibinfo {year} {2019})}\BibitemShut
  {NoStop}%
\bibitem [{\citenamefont {Lucas}(2014)}]{lucas2014ising}%
  \BibitemOpen
  \bibfield  {author} {\bibinfo {author} {\bibfnamefont {A.}~\bibnamefont
  {Lucas}},\ }\bibfield  {title} {\bibinfo {title} {Ising formulations of many
  np problems},\ }\href@noop {} {\bibfield  {journal} {\bibinfo  {journal}
  {Frontiers in Physics}\ }\textbf {\bibinfo {volume} {2}},\ \bibinfo {pages}
  {5} (\bibinfo {year} {2014})}\BibitemShut {NoStop}%
\bibitem [{\citenamefont {Choi}(2010)}]{choi2010adiabatic}%
  \BibitemOpen
  \bibfield  {author} {\bibinfo {author} {\bibfnamefont {V.}~\bibnamefont
  {Choi}},\ }\bibfield  {title} {\bibinfo {title} {Adiabatic quantum algorithms
  for the np-complete maximum-weight independent set, exact cover and 3sat
  problems},\ }\href@noop {} {\bibfield  {journal} {\bibinfo  {journal} {arXiv
  preprint arXiv:1004.2226}\ } (\bibinfo {year} {2010})}\BibitemShut {NoStop}%
\bibitem [{\citenamefont {Burnett}\ \emph {et~al.}(2019)\citenamefont
  {Burnett}, \citenamefont {Bengtsson}, \citenamefont {Scigliuzzo},
  \citenamefont {Niepce}, \citenamefont {Kudra}, \citenamefont {Delsing},\ and\
  \citenamefont {Bylander}}]{burnett2019decoherence}%
  \BibitemOpen
  \bibfield  {author} {\bibinfo {author} {\bibfnamefont {J.~J.}\ \bibnamefont
  {Burnett}}, \bibinfo {author} {\bibfnamefont {A.}~\bibnamefont {Bengtsson}},
  \bibinfo {author} {\bibfnamefont {M.}~\bibnamefont {Scigliuzzo}}, \bibinfo
  {author} {\bibfnamefont {D.}~\bibnamefont {Niepce}}, \bibinfo {author}
  {\bibfnamefont {M.}~\bibnamefont {Kudra}}, \bibinfo {author} {\bibfnamefont
  {P.}~\bibnamefont {Delsing}},\ and\ \bibinfo {author} {\bibfnamefont
  {J.}~\bibnamefont {Bylander}},\ }\bibfield  {title} {\bibinfo {title}
  {Decoherence benchmarking of superconducting qubits},\ }\href@noop {}
  {\bibfield  {journal} {\bibinfo  {journal} {npj Quantum Information}\
  }\textbf {\bibinfo {volume} {5}},\ \bibinfo {pages} {9} (\bibinfo {year}
  {2019})}\BibitemShut {NoStop}%
\bibitem [{\citenamefont {McKay}\ \emph {et~al.}(2016)\citenamefont {McKay},
  \citenamefont {Filipp}, \citenamefont {Mezzacapo}, \citenamefont {Magesan},
  \citenamefont {Chow},\ and\ \citenamefont {Gambetta}}]{McKayUniversal}%
  \BibitemOpen
  \bibfield  {author} {\bibinfo {author} {\bibfnamefont {D.~C.}\ \bibnamefont
  {McKay}}, \bibinfo {author} {\bibfnamefont {S.}~\bibnamefont {Filipp}},
  \bibinfo {author} {\bibfnamefont {A.}~\bibnamefont {Mezzacapo}}, \bibinfo
  {author} {\bibfnamefont {E.}~\bibnamefont {Magesan}}, \bibinfo {author}
  {\bibfnamefont {J.~M.}\ \bibnamefont {Chow}},\ and\ \bibinfo {author}
  {\bibfnamefont {J.~M.}\ \bibnamefont {Gambetta}},\ }\bibfield  {title}
  {\bibinfo {title} {Universal gate for fixed-frequency qubits via a tunable
  bus},\ }\href@noop {} {\bibfield  {journal} {\bibinfo  {journal} {Physical
  Review Applied}\ }\textbf {\bibinfo {volume} {6}},\ \bibinfo {pages} {064007}
  (\bibinfo {year} {2016})}\BibitemShut {NoStop}%
\bibitem [{\citenamefont {Caldwell}\ \emph {et~al.}(2018)\citenamefont
  {Caldwell}, \citenamefont {Didier}, \citenamefont {Ryan}, \citenamefont
  {Sete}, \citenamefont {Hudson}, \citenamefont {Karalekas}, \citenamefont
  {Manenti}, \citenamefont {da~Silva}, \citenamefont {Sinclair}, \citenamefont
  {Acala} \emph {et~al.}}]{caldwell2018parametrically}%
  \BibitemOpen
  \bibfield  {author} {\bibinfo {author} {\bibfnamefont {S.}~\bibnamefont
  {Caldwell}}, \bibinfo {author} {\bibfnamefont {N.}~\bibnamefont {Didier}},
  \bibinfo {author} {\bibfnamefont {C.}~\bibnamefont {Ryan}}, \bibinfo {author}
  {\bibfnamefont {E.}~\bibnamefont {Sete}}, \bibinfo {author} {\bibfnamefont
  {A.}~\bibnamefont {Hudson}}, \bibinfo {author} {\bibfnamefont
  {P.}~\bibnamefont {Karalekas}}, \bibinfo {author} {\bibfnamefont
  {R.}~\bibnamefont {Manenti}}, \bibinfo {author} {\bibfnamefont
  {M.}~\bibnamefont {da~Silva}}, \bibinfo {author} {\bibfnamefont
  {R.}~\bibnamefont {Sinclair}}, \bibinfo {author} {\bibfnamefont
  {E.}~\bibnamefont {Acala}}, \emph {et~al.},\ }\bibfield  {title} {\bibinfo
  {title} {Parametrically activated entangling gates using transmon qubits},\
  }\href@noop {} {\bibfield  {journal} {\bibinfo  {journal} {Physical Review
  Applied}\ }\textbf {\bibinfo {volume} {10}},\ \bibinfo {pages} {034050}
  (\bibinfo {year} {2018})}\BibitemShut {NoStop}%
\bibitem [{\citenamefont {McKay}\ \emph {et~al.}(2017)\citenamefont {McKay},
  \citenamefont {Wood}, \citenamefont {Sheldon}, \citenamefont {Chow},\ and\
  \citenamefont {Gambetta}}]{mckay2017efficient}%
  \BibitemOpen
  \bibfield  {author} {\bibinfo {author} {\bibfnamefont {D.~C.}\ \bibnamefont
  {McKay}}, \bibinfo {author} {\bibfnamefont {C.~J.}\ \bibnamefont {Wood}},
  \bibinfo {author} {\bibfnamefont {S.}~\bibnamefont {Sheldon}}, \bibinfo
  {author} {\bibfnamefont {J.~M.}\ \bibnamefont {Chow}},\ and\ \bibinfo
  {author} {\bibfnamefont {J.~M.}\ \bibnamefont {Gambetta}},\ }\bibfield
  {title} {\bibinfo {title} {Efficient z gates for quantum computing},\
  }\href@noop {} {\bibfield  {journal} {\bibinfo  {journal} {Physical Review
  A}\ }\textbf {\bibinfo {volume} {96}},\ \bibinfo {pages} {022330} (\bibinfo
  {year} {2017})}\BibitemShut {NoStop}%
\bibitem [{\citenamefont {C{\'o}rcoles}\ \emph {et~al.}(2013)\citenamefont
  {C{\'o}rcoles}, \citenamefont {Gambetta}, \citenamefont {Chow}, \citenamefont
  {Smolin}, \citenamefont {Ware}, \citenamefont {Strand}, \citenamefont
  {Plourde},\ and\ \citenamefont {Steffen}}]{corcoles2013process}%
  \BibitemOpen
  \bibfield  {author} {\bibinfo {author} {\bibfnamefont {A.~D.}\ \bibnamefont
  {C{\'o}rcoles}}, \bibinfo {author} {\bibfnamefont {J.~M.}\ \bibnamefont
  {Gambetta}}, \bibinfo {author} {\bibfnamefont {J.~M.}\ \bibnamefont {Chow}},
  \bibinfo {author} {\bibfnamefont {J.~A.}\ \bibnamefont {Smolin}}, \bibinfo
  {author} {\bibfnamefont {M.}~\bibnamefont {Ware}}, \bibinfo {author}
  {\bibfnamefont {J.}~\bibnamefont {Strand}}, \bibinfo {author} {\bibfnamefont
  {B.~L.~T.}\ \bibnamefont {Plourde}},\ and\ \bibinfo {author} {\bibfnamefont
  {M.}~\bibnamefont {Steffen}},\ }\bibfield  {title} {\bibinfo {title} {Process
  verification of two-qubit quantum gates by randomized benchmarking},\
  }\href@noop {} {\bibfield  {journal} {\bibinfo  {journal} {Physical Review
  A}\ }\textbf {\bibinfo {volume} {87}},\ \bibinfo {pages} {030301(R)}
  (\bibinfo {year} {2013})}\BibitemShut {NoStop}%
\bibitem [{\citenamefont {Michielsen}\ \emph {et~al.}(2017)\citenamefont
  {Michielsen}, \citenamefont {Nocon}, \citenamefont {Willsch}, \citenamefont
  {Jin}, \citenamefont {Lippert},\ and\ \citenamefont
  {De~Raedt}}]{michielsen2017benchmarking}%
  \BibitemOpen
  \bibfield  {author} {\bibinfo {author} {\bibfnamefont {K.}~\bibnamefont
  {Michielsen}}, \bibinfo {author} {\bibfnamefont {M.}~\bibnamefont {Nocon}},
  \bibinfo {author} {\bibfnamefont {D.}~\bibnamefont {Willsch}}, \bibinfo
  {author} {\bibfnamefont {F.}~\bibnamefont {Jin}}, \bibinfo {author}
  {\bibfnamefont {T.}~\bibnamefont {Lippert}},\ and\ \bibinfo {author}
  {\bibfnamefont {H.}~\bibnamefont {De~Raedt}},\ }\bibfield  {title} {\bibinfo
  {title} {Benchmarking gate-based quantum computers},\ }\href@noop {}
  {\bibfield  {journal} {\bibinfo  {journal} {Computer Physics Communications}\
  }\textbf {\bibinfo {volume} {220}},\ \bibinfo {pages} {44} (\bibinfo {year}
  {2017})}\BibitemShut {NoStop}%
\bibitem [{\citenamefont {Kjaergaard}\ \emph
  {et~al.}(2020{\natexlab{b}})\citenamefont {Kjaergaard}, \citenamefont
  {Schwartz}, \citenamefont {Greene}, \citenamefont {Samach}, \citenamefont
  {Bengtsson}, \citenamefont {O'Keeffe}, \citenamefont {McNally}, \citenamefont
  {Braum{\"u}ller}, \citenamefont {Kim}, \citenamefont {Krantz} \emph
  {et~al.}}]{kjaergaard2020quantum}%
  \BibitemOpen
  \bibfield  {author} {\bibinfo {author} {\bibfnamefont {M.}~\bibnamefont
  {Kjaergaard}}, \bibinfo {author} {\bibfnamefont {M.}~\bibnamefont
  {Schwartz}}, \bibinfo {author} {\bibfnamefont {A.}~\bibnamefont {Greene}},
  \bibinfo {author} {\bibfnamefont {G.}~\bibnamefont {Samach}}, \bibinfo
  {author} {\bibfnamefont {A.}~\bibnamefont {Bengtsson}}, \bibinfo {author}
  {\bibfnamefont {M.}~\bibnamefont {O'Keeffe}}, \bibinfo {author}
  {\bibfnamefont {C.}~\bibnamefont {McNally}}, \bibinfo {author} {\bibfnamefont
  {J.}~\bibnamefont {Braum{\"u}ller}}, \bibinfo {author} {\bibfnamefont
  {D.}~\bibnamefont {Kim}}, \bibinfo {author} {\bibfnamefont {P.}~\bibnamefont
  {Krantz}}, \emph {et~al.},\ }\bibfield  {title} {\bibinfo {title} {A quantum
  instruction set implemented on a superconducting quantum processor},\
  }\href@noop {} {\bibfield  {journal} {\bibinfo  {journal} {arXiv preprint
  arXiv:2001.08838}\ } (\bibinfo {year} {2020}{\natexlab{b}})}\BibitemShut
  {NoStop}%
\bibitem [{\citenamefont {Chow}\ \emph {et~al.}(2010)\citenamefont {Chow},
  \citenamefont {DiCarlo}, \citenamefont {Gambetta}, \citenamefont
  {Nunnenkamp}, \citenamefont {Bishop}, \citenamefont {Frunzio}, \citenamefont
  {Devoret}, \citenamefont {Girvin},\ and\ \citenamefont
  {Schoelkopf}}]{chow2010detecting}%
  \BibitemOpen
  \bibfield  {author} {\bibinfo {author} {\bibfnamefont {J.~M.}\ \bibnamefont
  {Chow}}, \bibinfo {author} {\bibfnamefont {L.}~\bibnamefont {DiCarlo}},
  \bibinfo {author} {\bibfnamefont {J.~M.}\ \bibnamefont {Gambetta}}, \bibinfo
  {author} {\bibfnamefont {A.}~\bibnamefont {Nunnenkamp}}, \bibinfo {author}
  {\bibfnamefont {L.~S.}\ \bibnamefont {Bishop}}, \bibinfo {author}
  {\bibfnamefont {L.}~\bibnamefont {Frunzio}}, \bibinfo {author} {\bibfnamefont
  {M.~H.}\ \bibnamefont {Devoret}}, \bibinfo {author} {\bibfnamefont {S.~M.}\
  \bibnamefont {Girvin}},\ and\ \bibinfo {author} {\bibfnamefont {R.~J.}\
  \bibnamefont {Schoelkopf}},\ }\bibfield  {title} {\bibinfo {title} {Detecting
  highly entangled states with a joint qubit readout},\ }\href@noop {}
  {\bibfield  {journal} {\bibinfo  {journal} {Physical Review A}\ }\textbf
  {\bibinfo {volume} {81}},\ \bibinfo {pages} {062325} (\bibinfo {year}
  {2010})}\BibitemShut {NoStop}%
\end{thebibliography}%
\end{document}